\documentclass[letterpaper]{article} % DO NOT CHANGE THIS
\pdfoutput=1
\usepackage{aaai23}
\usepackage{times}  % DO NOT CHANGE THIS
\usepackage{helvet}  % DO NOT CHANGE THIS
\usepackage{courier}  % DO NOT CHANGE THIS
\usepackage[hyphens]{url}  % DO NOT CHANGE THIS
\usepackage{graphicx} % DO NOT CHANGE THIS
\usepackage{natbib}  % DO NOT CHANGE THIS AND DO NOT ADD ANY OPTIONS TO IT
\usepackage{caption} % DO NOT CHANGE THIS AND DO NOT ADD ANY OPTIONS TO IT
\usepackage{subcaption}     % subfigures (mine)
\usepackage{amsmath}        % math (mine)
\usepackage{amssymb}        % math (mine)
\usepackage{booktabs}       % professional-quality tables
\usepackage{multirow}
\usepackage{algorithm}
\usepackage{algorithmic}

\usepackage{newfloat}
\usepackage{listings}
\DeclareCaptionStyle{ruled}{labelfont=normalfont,labelsep=colon,strut=off} % DO NOT CHANGE THIS
\floatstyle{ruled}
\newfloat{listing}{tb}{lst}{}
\floatname{listing}{Listing}
\title{DELAD: Deep Landweber-guided deconvolution with Hessian and sparse prior}
\author{
    Tomas Chobola\textsuperscript{\rm 1,2},
    Anton Theileis\textsuperscript{\rm 3},
    Jan Taucher\textsuperscript{\rm 3},
    Tingying Peng\textsuperscript{\rm 1,2}
}
\affiliations{
    \textsuperscript{\rm 1}Technical University of Munich, Munich, Germany\\
    \textsuperscript{\rm 2}Helmholtz AI, Neuherberg, Germany\\
    \textsuperscript{\rm 3}GEOMAR Helmholtz Centre for Ocean Research Kiel, Kiel, Germany
}

\usepackage{bibentry}
\begin{document}

\nocopyright % <------ NEEDS TO BE DELETED FOR PUBLISHING

\maketitle

\begin{abstract}
We present a model for non-blind image deconvolution that incorporates the classic iterative method into a deep learning application. Instead of using large over-parameterised generative networks to create sharp picture representations, we build our network based on the iterative Landweber deconvolution algorithm, which is integrated with trainable convolutional layers to enhance the recovered image structures and details. Additional to the data fidelity term, we also add Hessian and sparse constraints as regularization terms to improve the image reconstruction quality. Our proposed model is \textit{self-supervised} and converges to a solution based purely on the input blurred image and respective blur kernel without the requirement of any pre-training. We evaluate our technique using standard computer vision benchmarking datasets as well as real microscope images obtained by our enhanced depth-of-field (EDOF) underwater microscope, demonstrating the capabilities of our model in a real-world application. The quantitative results demonstrate that our approach is competitive with state-of-the-art non-blind image deblurring methods despite having a fraction of the parameters and not being pre-trained, demonstrating the efficiency and efficacy of embedding a classic deconvolution approach inside a deep network.
\end{abstract}

% ------------------------------------------------------------------------
\section{Introduction}

Image blurring can be mathematically expressed as follows: 
\begin{equation}\label{eq:conv}
    % \mathbf{y}=\mathbf{x}\otimes\mathbf{k}+\mathbf{n},
    \mathbf{y}=\mathbf{H}\mathbf{x}+\mathbf{b},
\end{equation}
where $\mathbf{y}$ is the blurred image obtained by convolving a sharp image $\mathbf{x}$ with $\mathbf{H}$, which is a square matrix that approximates the convolution with the PSF, and $\mathbf{b}$ represents the measurement error. Image deconvolution aims to reverse the blurring process and reconstruct the sharp image. Existing image deconvolution methods can be generally divided into \textbf{non-blind deconvolution} and \textbf{blind deconvolution}, depending on whether $\mathbf{H}$ is known or not.

The recovery of $\mathbf{x}$ from $\mathbf{y}$ is an ill-posed problem \cite{illposed}. In order to obtain an estimation with feasible accuracy, we need to rely on prior information about the image space. The quality of the reconstructed image is typically measured with a cost function as follows,
\begin{equation}
    \label{eq:deconv}
    \mathcal{J}(\mathbf{x})=\mathcal{J}_1(\mathbf{x}) + \lambda\mathcal{J}_2(\mathbf{x}),
\end{equation}
where $\mathcal{J}_1(\cdot)$ appraises the estimate with respect to the measurements, and $\mathcal{J}_2(\cdot)$ penalises the absence of desired properties. The parameter $\lambda$ balances the importance of the second term in relation to the first term.

% ------------------------------------------------------------------------
\section{Related work}

\paragraph{Non-blind deconvolution.} Early methods for restoring a sharp image given its blurred representation and respective blur kernel include the Wiener filter \cite{wiener1949filter}, which imposes a Gaussian assumption on the noise, the Richardson-Lucy algorithm \cite{Richardson:72}, which assumes the noise follows a Poisson distribution, and the Landweber iteration \cite{landweber}. Additionally, multiple methods propose to use an iterative optimization based on MAP estimations \cite{levin2007image, krishnan2009lap}, but differ in their employed priors (e.g., global image or local patch-based). Popular priors include total variation \cite{wang2008tv} or hyper-Laplacian priors \cite{krishnan2009lap} that help with constraining the solution space and improving the estimation. 

Recent deep learning methods use a U-shape network to perform the deconvolution in the image space and to reconstruct the sharp image \cite{Zhang_2017_CVPR, schuler2013, IRCNN}. Particularly, an interesting approach was proposed to integrate the classic Wiener filter into a deep learning application \cite{dwdn}. Instead of applying the Wiener filter in standard image space, \cite{dwdn} performs Wiener filtering on deep features extracted by CNNs, which produces less noisy results compared to pure network-based deconvolution algorithms. The Richardson-Lucy algorithm \cite{Richardson:72} was also combined with deep learning into a model that when pre-trained showed improved deconvolution performance \cite{rl_net}. Recently, Zhao \textit{et al.} \cite{nature_sparse} took the advantage of a priori knowledge about the sparsity and continuity of biological structures to adapt the Richardson-Lucy method \cite{Richardson:72} to increase the resolution of live-cell microscopes nearly twofold.

\paragraph{Blind deconvolution.} Recovering the clear picture only from its blurred representation without knowing the blur kernel is even more challenging than non-blind deconvolution. Many previous blind deconvolution methods rely on a maximum a posterior (MAP) framework and Bayesian methods \cite{fergus2006removing, levin, xu2010two, pan2016blind}. Usually, the sharp image and the kernel are estimated in an alternating fashion, with further constraints such as L1/L2 norm \cite{krishnan2011blind} or patch-based prior \cite{michaeli2014blind} to find the optimal solution. It has also been shown that studying the problem in the frequency domain can lead to high-quality kernel estimation \cite{pan2019phase}. The non-blind Richardson-Lucy algorithm \cite{Richardson:72} has been successfully adapted to estimate the kernel as well, turning it into a blind method \cite{blind_rl}. This observation has been used by Agarwal \textit{et al.} \cite{deep-url} to replace the initial sharp image and kernel estimates with learnable weights resulting in a deep unfolded Richardson-Lucy framework that outperforms the classic approach.

\paragraph{Our contribution.} The contributions of this paper can be summarised as follows: \textit{(i)} We propose DELAD, a self-supervised approach for non-blind deconvolution embedding the Landweber iteration, a classic method for image deconvolution, into a deep learning model, omitting the need for over-parameterised generative networks which results in a lightweight and effective model. \textit{(ii)} We integrate the continuity prior as a form of Hessian regularisation into the loss function reducing the noise in the resulting sharp image while preserving the underlying image structures. \textit{(iii)} We apply our proposed DELAD to deconvolve real microscopic plankton images captured by our underwater camera with an extended depth-of-field (EDOF) lens, thereby removing the blurring effects caused by out-of-focus image slices. The source code is publicly available at https://github.com/ctom2/delad.

\begin{figure*}[h!]
    \centering
    \includegraphics[width=.75\linewidth]{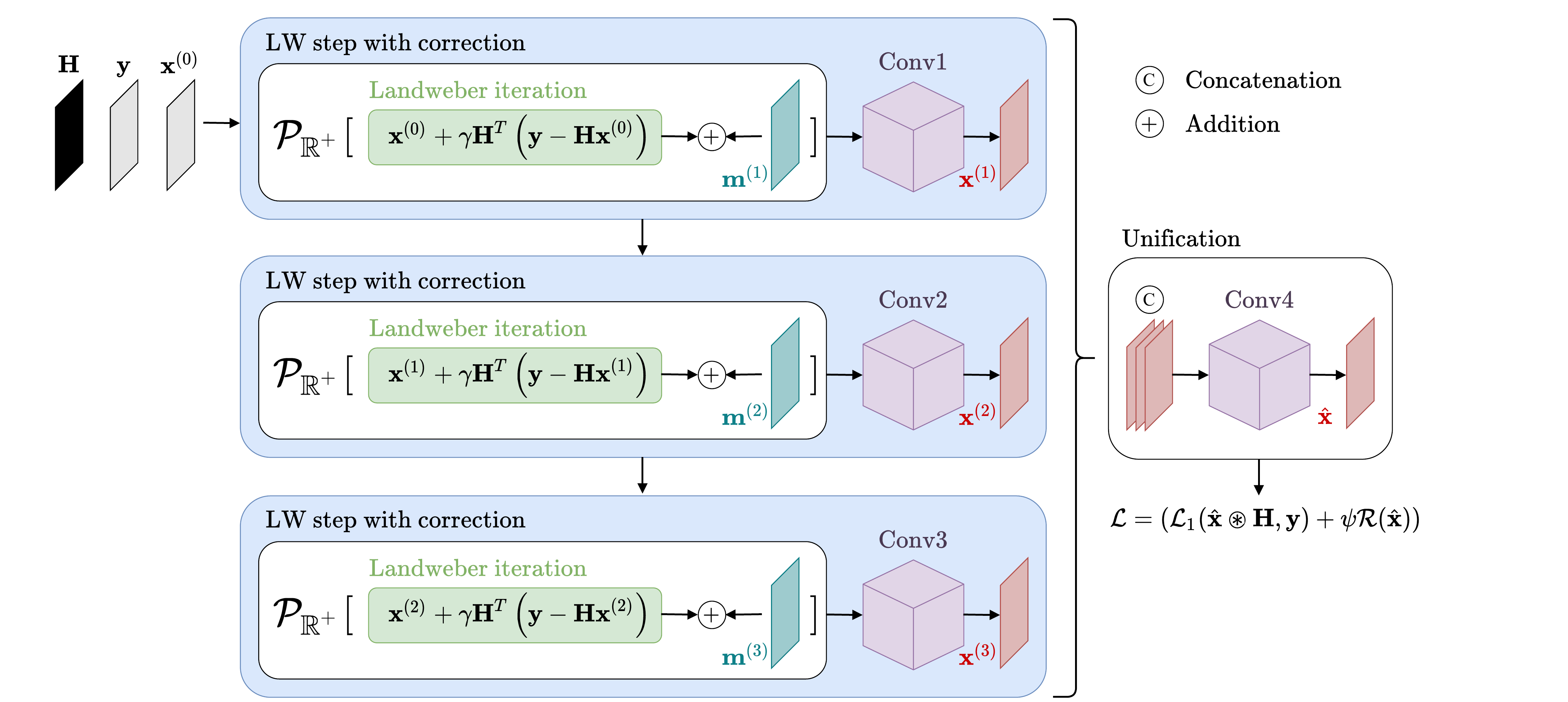}
    \caption{The architecture of the proposed non-blind deconvolution model embedding the Landweber iteration. The terms $\mathbf{x}^{(0)},\mathbf{m}^{(1)},\mathbf{m}^{(2)},\mathbf{m}^{(3)}$ together with kernels in the convolutional layers are updated during the optimisation of the model.}
    \label{fig:architecture}
\end{figure*}

% ------------------------------------------------------------------------
\section{Methodology}

We propose a combination of a classic non-blind iterative deconvolution algorithm and deep learning. The objective of the method is to restore a representation of the sharp image based purely on the blurry input image and the appropriate kernel in a \textit{self-supervised} manner. By embedding Landweber iteration \cite{landweber} into the algorithm we omit the need for using an extensively large neural network to produce the sharp image. We adopt Landweber iteration \cite{landweber} for its simplicity \cite{simple} over other iterative deconvolution algorithms such as Richardson-Lucy \cite{Richardson:72}.

\subsection{Landweber iteration}

To obtain the estimate of the sharp image $\mathbf{x}$ from a degraded measurement $\mathbf{y}$ and known kernel $\mathbf{H}$, following Equation \ref{eq:deconv}, the Landweber algorithm \cite{landweber, lw2} quantifies the quality of the estimate as
\begin{equation}
    \label{eq:lw_prior}
    \mathcal{J}_1=\lVert\mathbf{y} - \mathbf{Hx}\rVert_2^2.
\end{equation}
The term measures the residual using the squared L2 norm $\lVert\cdot\rVert_2$. By assuming $\lambda=0$ in Equation \ref{eq:deconv}, the minimiser of the expression becomes the least square estimator given by $\mathbf{x}=\mathbf{H}^\dagger\mathbf{y}$, where
\begin{equation}
    \mathbf{H}^\dagger=\lim\limits_{c\rightarrow 0, c>0}\left(\mathbf{H}^T\mathbf{H}+c\mathbf{I}\right)^{-1}\mathbf{H}^T
\end{equation}
is the pseudoinverse of $\mathbf{H}$. While $\mathbf{H}^\dagger\mathbf{y}$ can be computed in the frequency domain, where $\mathbf{H}^\dagger$ is diagonalised, the method is prone to noisy results as the matrix $\mathbf{H}^T\mathbf{H}$ can have small but non-zero eigenvalues. To prevent the noise amplification, Equation \ref{eq:lw_prior} is minimised in an iterative manner by constructing a sequence $(\mathbf{x}^{(n)})_{n\in\mathbb{N}}$ with an initial estimate $\mathbf{x}^{(0)}$ that converges to a minimiser of \ref{eq:lw_prior}. Applying gradient descent yields the following iterative formula,
\begin{equation}
    \label{eq:lw}
    \mathbf{x}^{(k+1)}=\mathbf{x}^{(k)}+\gamma\mathbf{H}^T(\mathbf{y}-\mathbf{Hx}^{(k)}),
\end{equation}
where $\gamma$ denotes the step size. 

\subsection{Deep learning-based Landweber-guided deconvolution with Hessian regularization}

To design a deep deconvolution architecture, we propose to embed an explicit Landweber iteration into the model (see Equation \ref{eq:lw}). We over-parameterize the iterations by substituting the initial estimate $\mathbf{x}^{(0)}$ with a matrix of learnable parameters and by adding correction terms $\mathbf{m}$, which are also learnable matrices of parameters of the same size as the degraded image $\mathbf{y}$, to the respective iterations. This Landweber step is projected to non-negative real numbers (denoted as $\mathcal{P}_{\mathbb{R}^+}$) and then passed through a respective convolutional layer $\mathcal{C}$. We can obtain the deep deconvolution step as follows,
\begin{align}
    \label{eq:deep_lw_step}
    \mathbf{x}^{(k+1)}&=\mathcal{C}^{(k+1)}\left\{\mathcal{P}_{\mathbb{R}^+}\left[\mathbf{W}\right]\right\},\\
    \mathbf{W}&=\mathbf{x}^{(k)}+\gamma\mathbf{H}^T(\mathbf{y}-\mathbf{Hx}^{(k)})+\mathbf{m}^{(k+1)}
\end{align}
The outputs of the convolutional layers $\mathcal{C}^{(k)}$ are then concatenated and passed to the final layer that unifies the intermediate deconvolution results $\mathbf{x}^{(k)}\;\forall k > 0$ into a single sharp image estimate $\hat{\mathbf{x}}$.

The terms $\mathbf{x}^{(0)},\mathbf{m}^{(1)},\mathbf{m}^{(2)},\mathbf{m}^{(3)}$ together with the kernels of convolutional layers $\mathcal{C}^{(k)}$ are updated in an iterative \textit{self-supervised} approach as the only information provided to the model is the degraded image $\mathbf{y}$ and the corresponding blur kernel $\mathbf{H}$. The parameters are optimised to minimise the following loss function,
\begin{equation}
    \label{eq:loss}
    \mathcal{L}=\mathcal{L}_1(\hat{\mathbf{x}}\circledast\mathbf{H},\mathbf{y}) + \psi\mathcal{R}(\hat{\mathbf{x}}),
\end{equation}
where $\mathcal{L}_1(\cdot)$ is the negative of the structural similarity index (SSIM) between the degraded image $\mathbf{y}$ and the representation of the sharp image $\hat{\mathbf{x}}$ convolved with the known kernel $\mathbf{H}$. The loss is moreover regularised with a Hessian prior in the image space
\begin{equation}
    \mathcal{R}(X)=\lVert{X_{xx}}\rVert_1 + \lVert{X_{yy}}\rVert_1 + 2\lVert{X_{xy}}\rVert_1,
\end{equation}
where $X_i$ denotes the second-order partial derivatives of $X$ along the $x$ and $y$ axes. The purpose of the regularisation is to reduce the noise artifacts arising from the deconvolution steps and to preserve the underlying image structure. The strength of the regularisation is controlled by the parameter $\psi$. Figure \ref{fig:architecture} illustrates the full model architecture used for obtaining the sharp image estimate $\hat{\mathbf{x}}$ with 3 deep Landweber-guided iterations and the loss function computation.

\subsection{Background estimation and sparsity prior for real  microscopy image deconvolution}
\label{subsection:background_sparsity}

For deconvolution of our EDOF microscopy images we perform pre-processing in a form of background removal ($\mathbf{b}$ from Equation \ref{eq:conv}). Moreover, we adapt the loss function by including sparsity as another prior additional to Hessian continuity prior. Sparsity prior has been shown to improve super-resolution microscopy image reconstruction \cite{nature_sparse}.

\paragraph{Background removal.} We follow the modified iterative wavelet transform method \cite{wavelet} described by Zhao \textit{et al.} \cite{nature_sparse}. We use the residual image arising from setting the values over the mean value of the original image $\mathbf{y}$ to zero to estimate the background as follows: \textit{(i)} The background is iteratively estimated from the lowest frequency wavelet bands related to the input image using 2D Daubechies-6 wavelet filters to decompose the signal up to the 7$^{\text{th}}$ level. \textit{(ii)} An inverse wavelet transform on the lowest band of the frequency information to the spatial domain is performed to prevent the unintended removal of important information. The result is then combined with $\sqrt{\mathbf{y}}/2$ into a single image whose pixels consist of the minimum of those two. \textit{(iii)} The output of the previous step is then used as the input in the next iteration. Following \cite{nature_sparse}, we set the number of iterations to 3 to estimate the background. The final estimated background is then subtracted from the original input image $\mathbf{y}$.

\paragraph{Sparsity.} Apart from continuity, another common feature in microscopy imaging is sparsity. We extend the loss function from Equation \ref{eq:loss} as follows,
\begin{equation}
    \label{eq:loss_sparse}
    \mathcal{L}=\mathcal{L}_1(\hat{\mathbf{x}}\circledast\mathbf{H},\mathbf{y}) + \psi_1\mathcal{R}(\hat{\mathbf{x}}) + \psi_2\lVert{\hat{\mathbf{x}}'}\rVert_1,
\end{equation}
where $\lVert{\cdot}\rVert_1$ denotes the L1 norm and $\hat{\mathbf{x}}'$ is the color inverted $\hat{\mathbf{x}}$. The reason for the inversion is that our real microscopy images consist of dark objects on light backgrounds, and therefore to enforce sparsity, we need to transform the input image where the background pixel values approach 0.

\begin{figure*}
    \captionsetup[subfigure]{font=scriptsize,labelfont=scriptsize}
    \centering
    \begin{subfigure}[t]{0.2\textwidth}
        \centering
        \includegraphics[width=\linewidth]{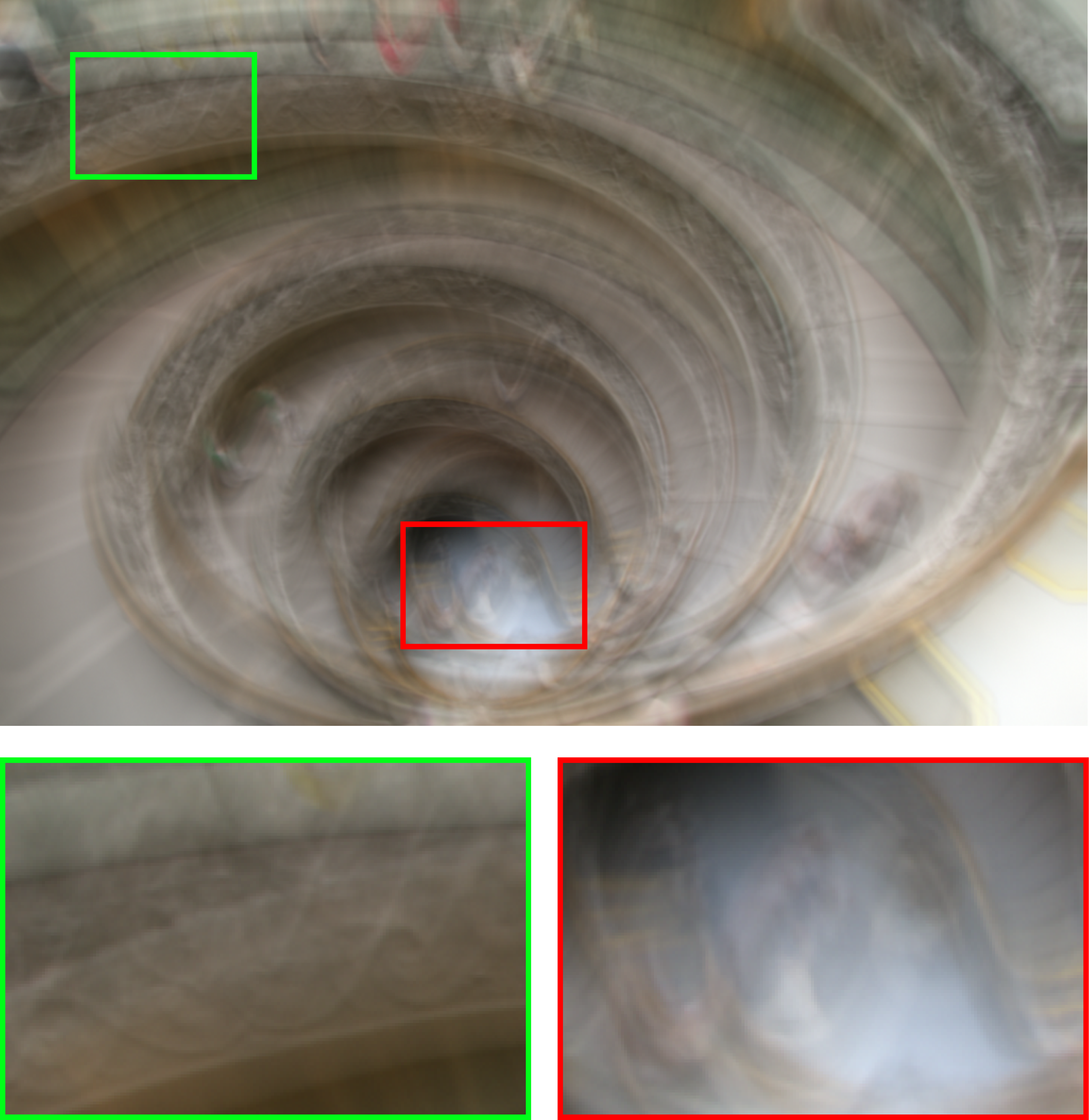}
        \caption{Blurry input}
        % \label{fig:kernel}
    \end{subfigure}
    \begin{subfigure}[t]{0.2\textwidth}
        \centering
        \includegraphics[width=\linewidth]{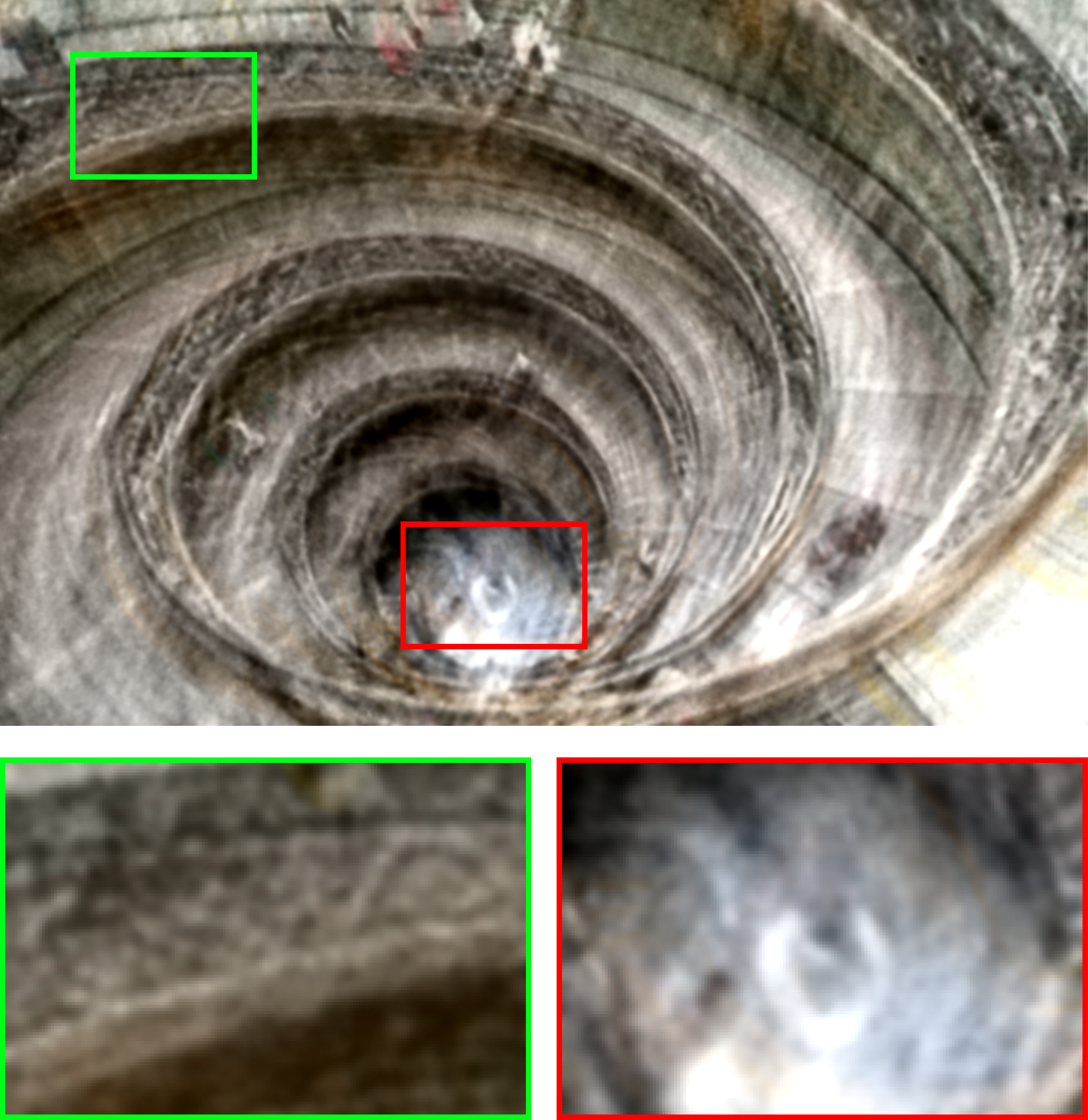}
        \caption{SelfDeblur$^\triangle$ \cite{selfdeblur}}
        \label{fig:lai_selfdeblur}
    \end{subfigure}
    \begin{subfigure}[t]{0.2\textwidth}
        \centering
        \includegraphics[width=\linewidth]{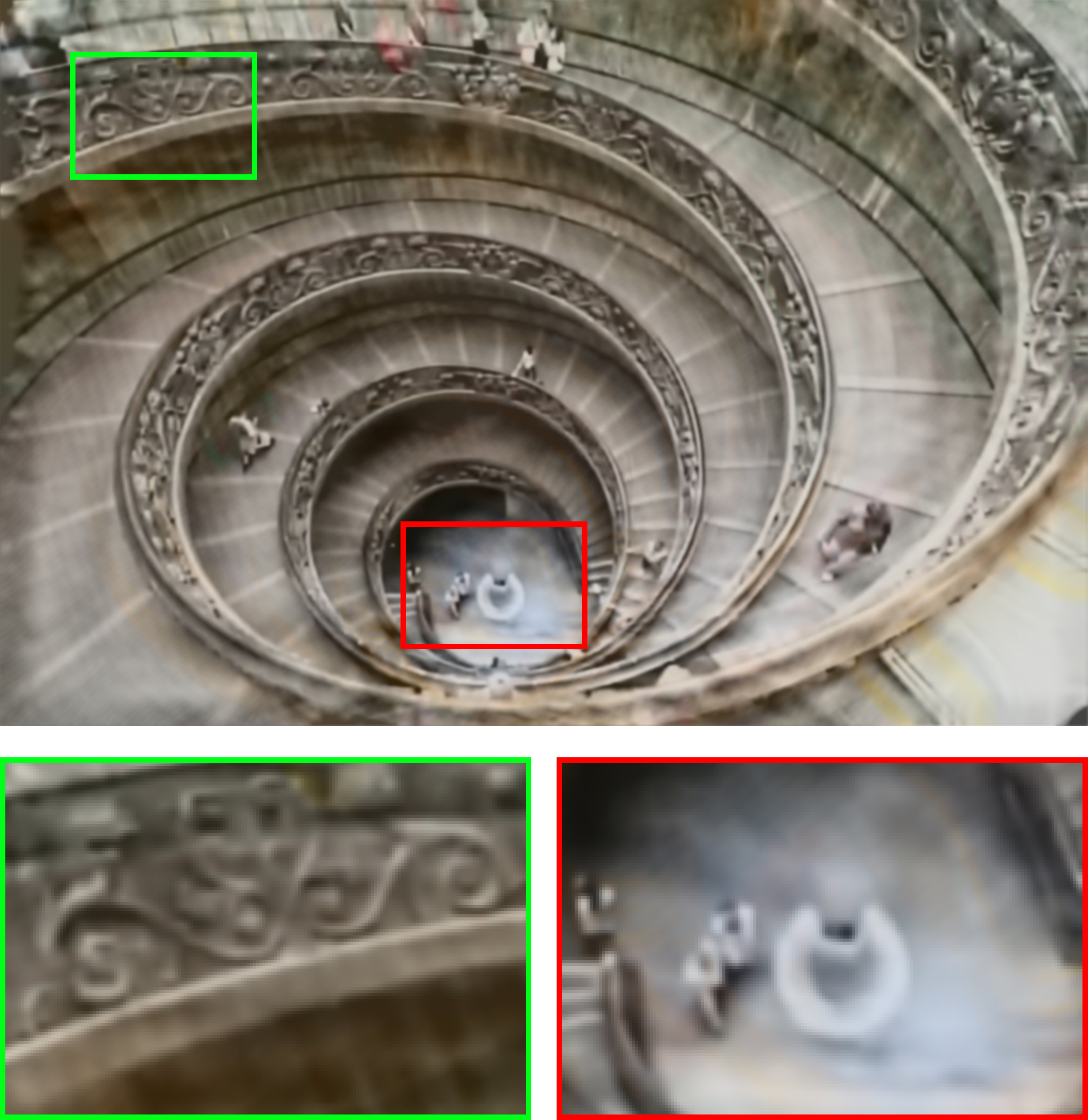}
        \caption{DELAD (ours)}
        \label{fig:lai_delad}
    \end{subfigure}
    \begin{subfigure}[t]{0.2\textwidth}
        \centering
        \includegraphics[width=\linewidth]{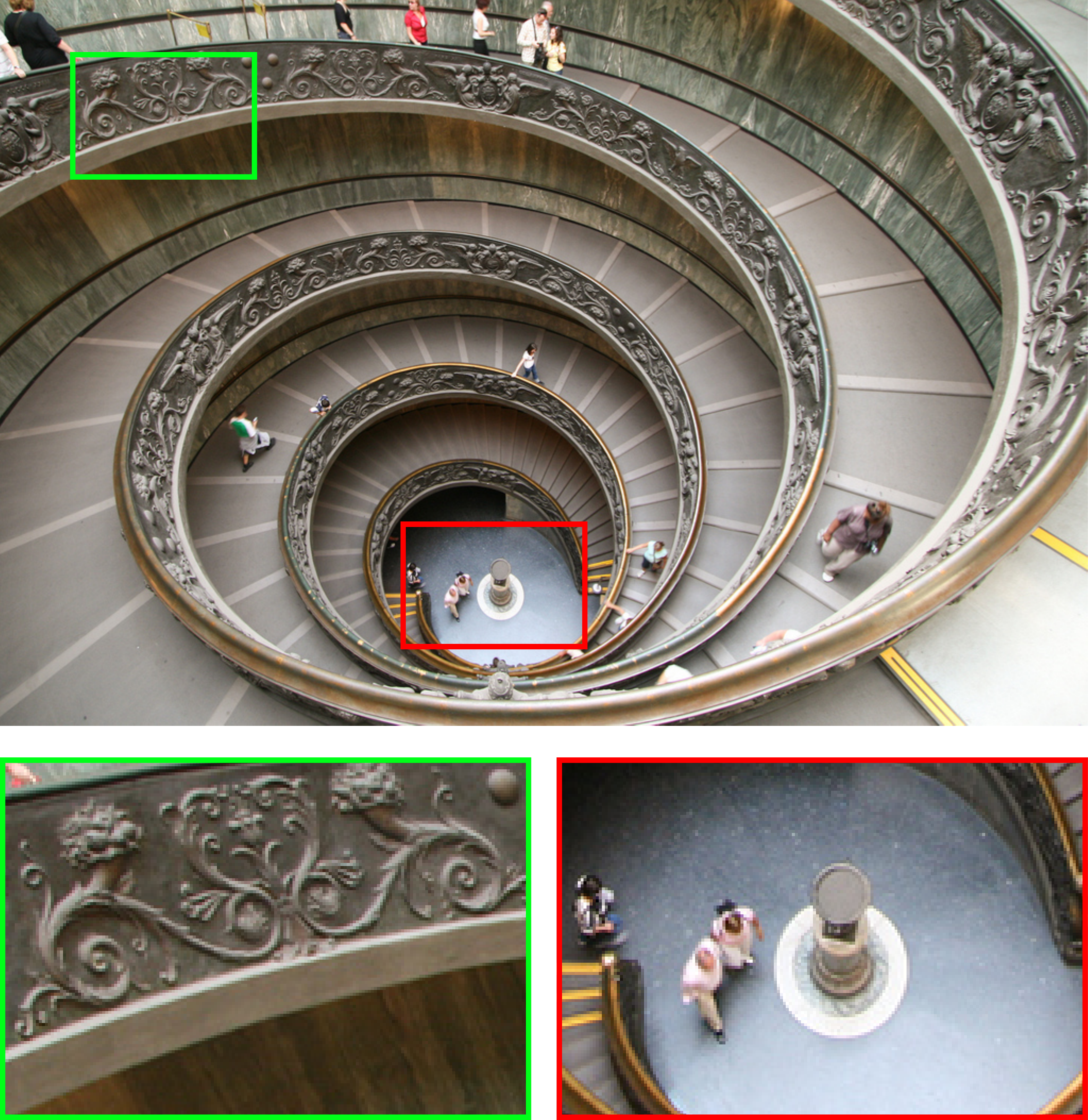}
        \caption{Ground truth}
        % \label{fig:gt}
    \end{subfigure}
    \caption{Overview of deconvolution outputs of two self-supervised methods. SelfDeblur$^\triangle$ result (Subfigure \ref{fig:lai_selfdeblur}) was obtained using a blind setting (no prior knowledge of the blur kernel), whereas DELAD result (Subfigure \ref{fig:lai_delad}) was obtained using a non-blind setting. The deconvolution with prior knowledge of the blur kernel allows for better reconstruction of the image details.}
\end{figure*}

% ------------------------------------------------------------------------
\section{Experimental results}

In this section, we investigate the performance of our proposed method and compare it to other state-of-the-art algorithms. We evaluate DELAD on a wide range of images including traditional computer vision quantitative datasets as well as simulated and real microscopy images.

\subsection{Implementation and datasets}
\label{subsection:implementation_and_data}

\paragraph{Implementation details.} To balance the computational complexity and efficacy the proposed model embeds three Landweber iterations. The learnable parts of the model are the initial matrix estimate $\mathbf{x}^{(0)}$, the correction terms for each iteration $\mathbf{m}^{(1)},\mathbf{m}^{(2)},\mathbf{m}^{(3)}$, and the kernels in the convolutional layers. The matrices are initialised with numbers from a uniform distribution in the interval $[0,1)$. We set the size of the kernels to be $3\times 3$ for every layer. In each Landweber iteration, we use the step size $\gamma=0.8$. Next, the projection of the iteration output to non-negative real numbers $\mathcal{P}_{\mathbb{R}^+}(\cdot)$ is done through the ReLU activation function. We use sigmoid as the activation function after each of the convolutional layers. 

The optimisation of the model is carried out by the RMSprop optimiser for 2000 epochs with the initial learning rate set to $0.05$ with a decaying factor of $0.2$ when reaching epochs 1000 and 1500. The strength of the continuity prior in the loss function (see Equation \ref{eq:loss}) is set to $\psi=1\mathrm{e}{-6}$.

\paragraph{Testing datasets.} To benchmark our proposed method and compare it to state-of-the-art methods we use the popular dataset Levin \textit{et al.} \cite{levin} comprising of four 255$\times$255 grayscale images and 8 motion blur kernels, resulting in a total of 32 blurred images. Further evaluation is done on the Lai \textit{et al.} \cite{lai_dataset} dataset that includes large color images and blur kernels of various sizes. Since the images are split into subsets \textsc{Manmade}, \textsc{Natural}, \textsc{People}, \textsc{Saturation}, and \textsc{Text}, we evaluate the methods separately for each set. Moreover, we perform an additional assessment on a dataset of synthetic microscopy images blurred with synthetic kernels following the evaluation of the unsupervised deconvolution algorithm proposed by Bredell \textit{et al.} \cite{wiener-dip}. The microscopy dataset takes four sharp synthetic images taken from the simulation of Schneider \textit{et al.} \cite{SCHNEIDER2015220} cropped to the size $255\times 255$. Each sharp image is blurred with one kernel from Levin \textit{et al.} \cite{levin} and three kernels expected in two-photon microscopy, resulting in a set of 16 blurred images.

% maybe move this EDOF section up, before the section about background removal etc.  then it is already clear, which device was used. 

\paragraph{Application on real "extended depth-of-field" images.} To prove the functionality of our proposed method in a scientific application, where image blur is a major limiting factor, we evaluate the performance on data obtained by an "extended depth-of-field" (EDOF) plankton imaging system. One major limitation of conventional optical microscopy is the trade-off between magnification and depth-of-field (DOF) which causes the DOF of an optical system to drastically reduce at higher magnifications. Consequently, small objects can only be examined in a very shallow focal plane, which is impractical for many applications that examine an extended sample volume. Image-based observations of plankton require both, a high level of image detail (magnification) and a large image volume (DOF) to collect representative samples of the plankton organisms in water \cite{LOMBARD}. Therefore, we developed a novel underwater imaging system to increase the sample volume while maintaining an optical resolution required to identify plankton ($>200\mu\text{m}$) by implementing an electrically tunable lens in a telecentric setup. The optical system axially integrates over all focal planes in the entire sample cube during one exposure time. This approach allows to substantially enhance the DOF and image volume while maintaining the desired optical resolution for the identification of plankton. However, as our EDOF image is captured in a single exposure, it contains both in-focus image slices (corresponding to the moving focal plane) and out-of-focus blurry slices (corresponding to those outside the focal range) that reduce the perceived resolution and contrast. Thus, the EDOF-based degradation in image quality makes subsequent data processing and analysis very problematic. So far, no solution exists to remove EDOF blur and improve the optical resolution of such system (see supplementary for the reconstruction results of EDOF images using SOTA deconvolution methods).

\begin{figure*}
    \captionsetup[subfigure]{font=scriptsize,labelfont=scriptsize,{justification=centering}}
    \centering
    \begin{subfigure}[t]{0.18\textwidth}
        \centering
        \includegraphics[width=\linewidth]{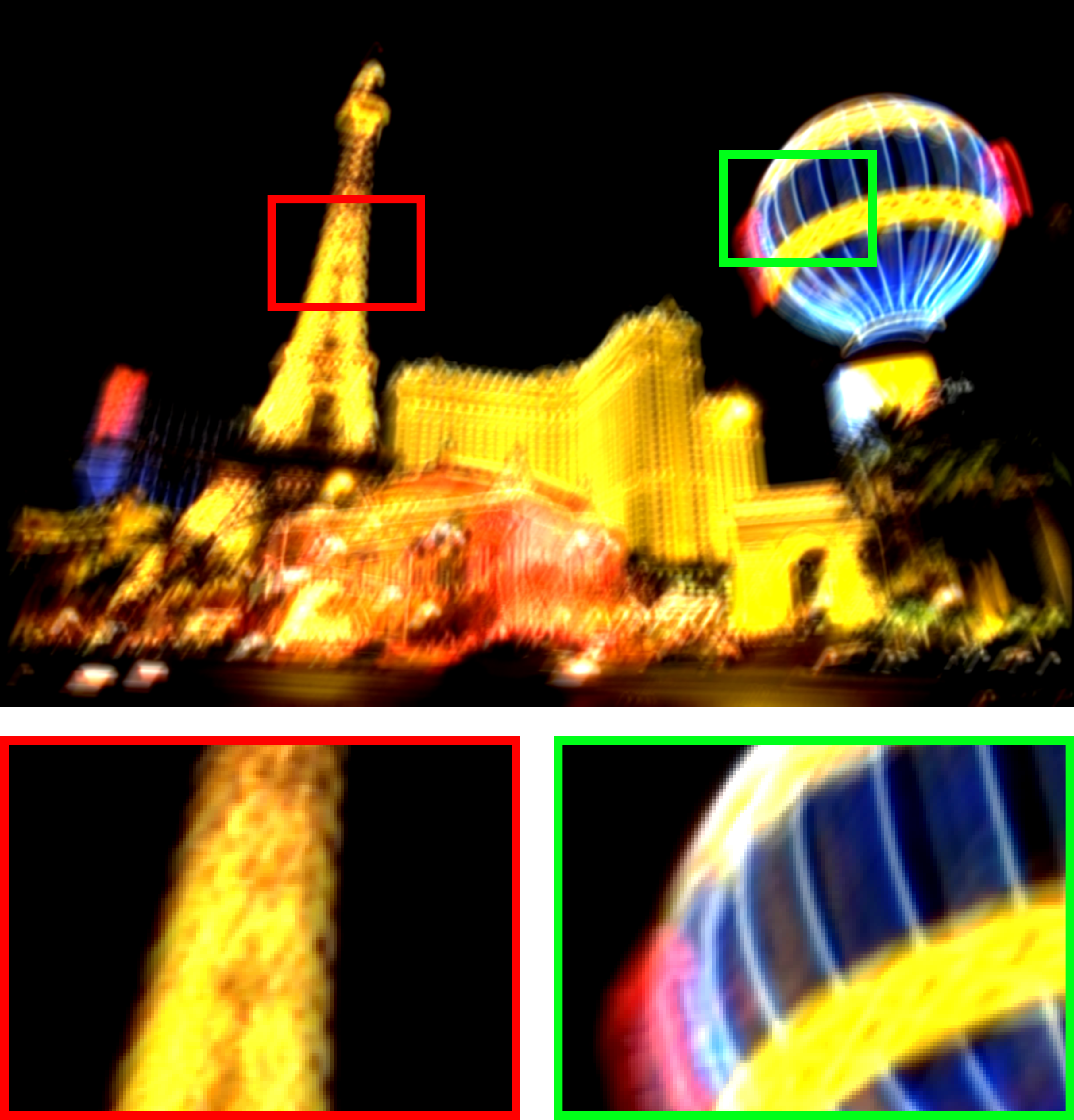}
        \caption{Blurry input}
    \end{subfigure}
    \begin{subfigure}[t]{0.18\textwidth}
        \centering
        \includegraphics[width=\linewidth]{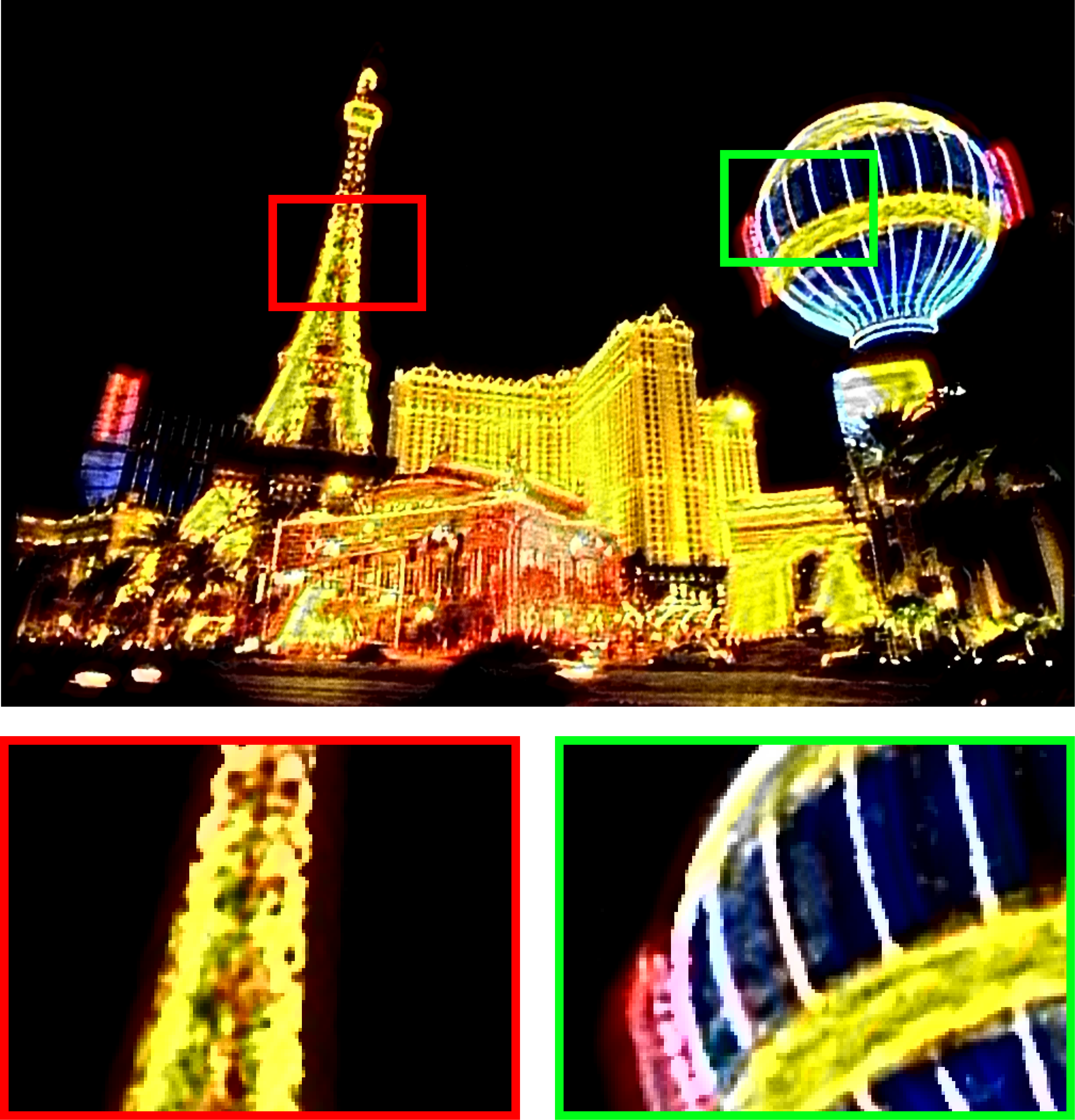}
        \caption{SelfDeblur$^*$\\\cite{selfdeblur}}
    \end{subfigure}
    \begin{subfigure}[t]{0.18\textwidth}
        \centering
        \includegraphics[width=\linewidth]{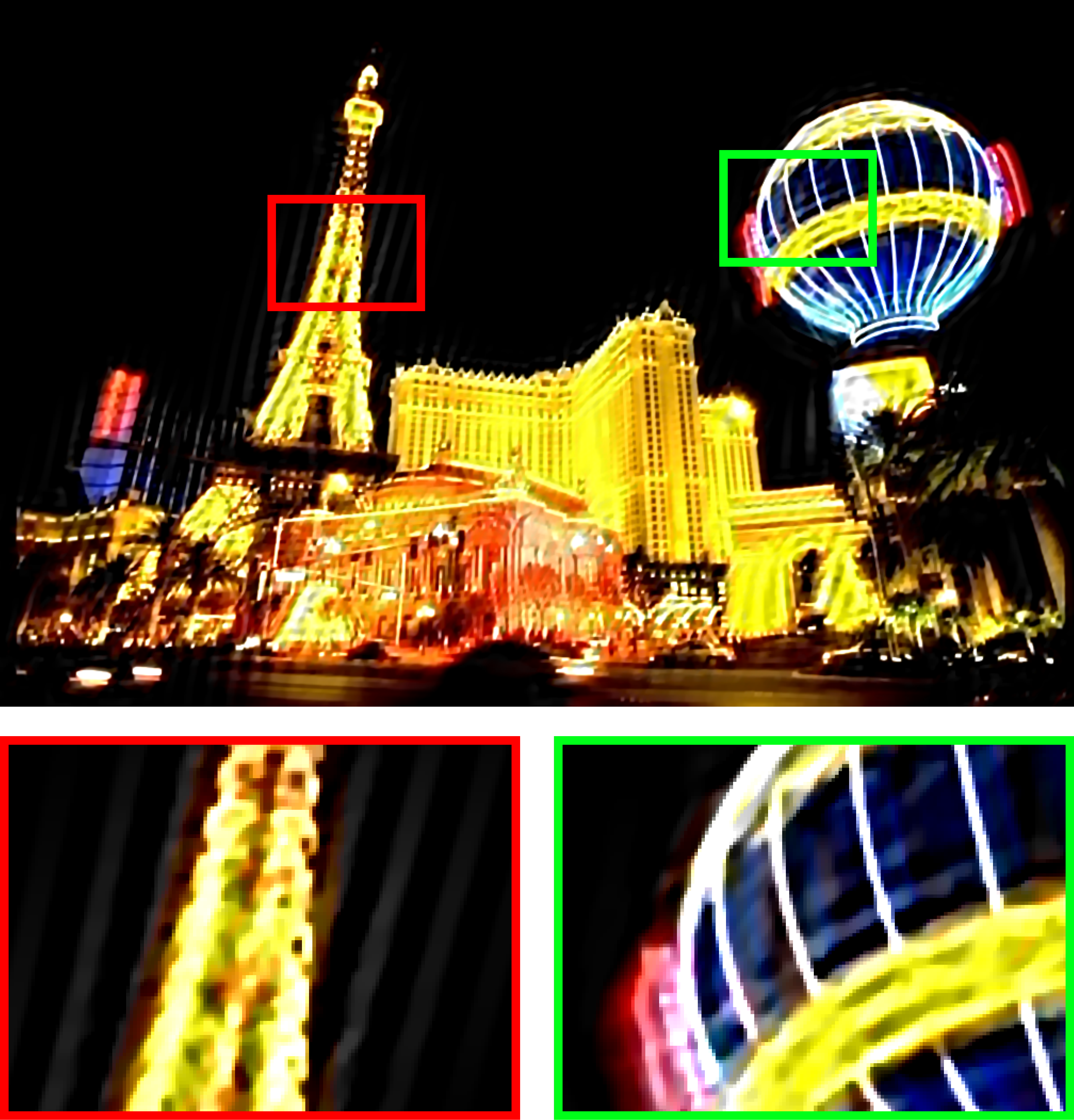}
        \caption{Pan-DCP\\\cite{pan2017deblurring}}
    \end{subfigure}
    \begin{subfigure}[t]{0.18\textwidth}
        \centering
        \includegraphics[width=\linewidth]{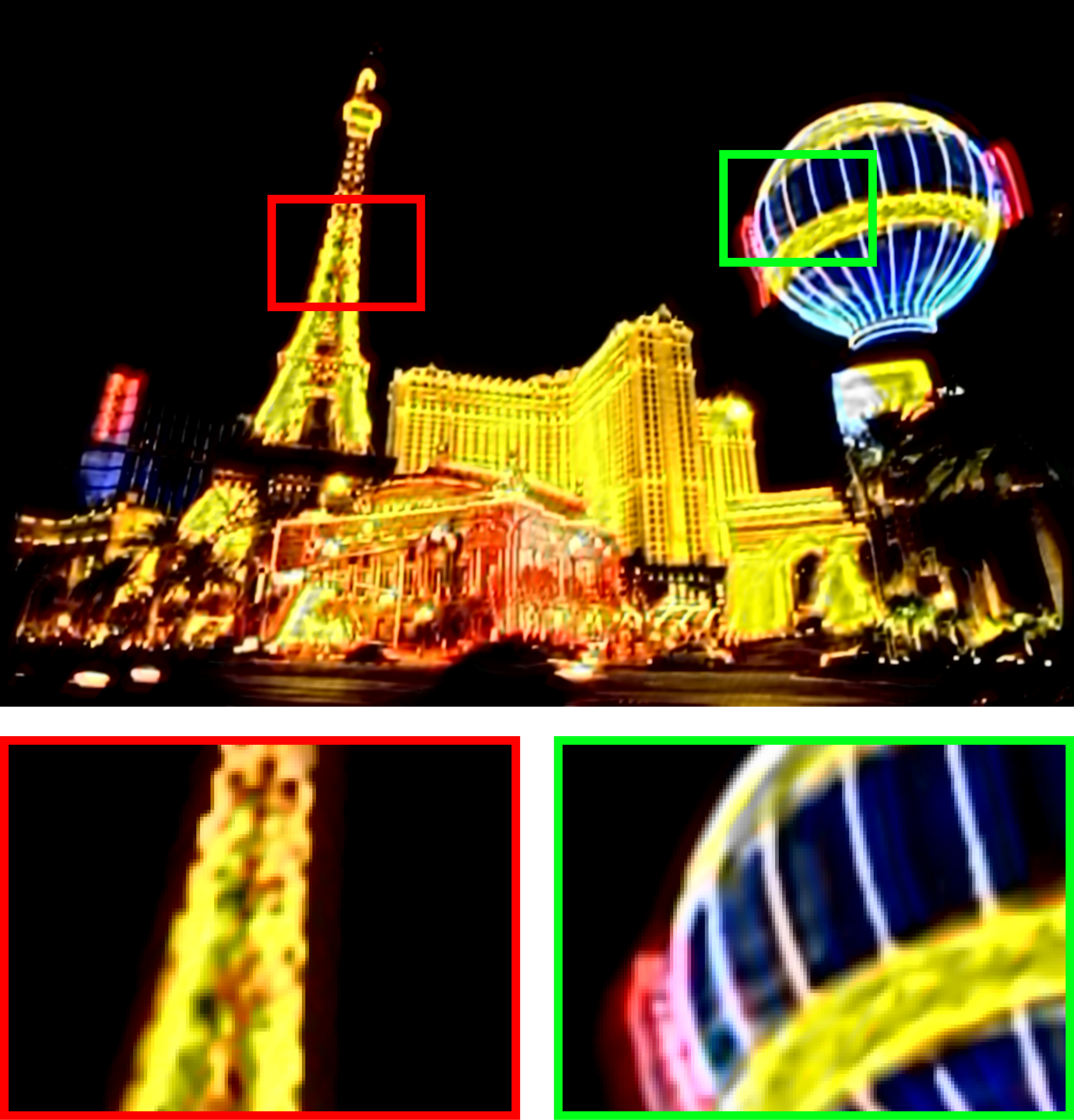}
        \caption{DELAD$^*$}
    \end{subfigure}
    \begin{subfigure}[t]{0.18\textwidth}
        \centering
        \includegraphics[width=\linewidth]{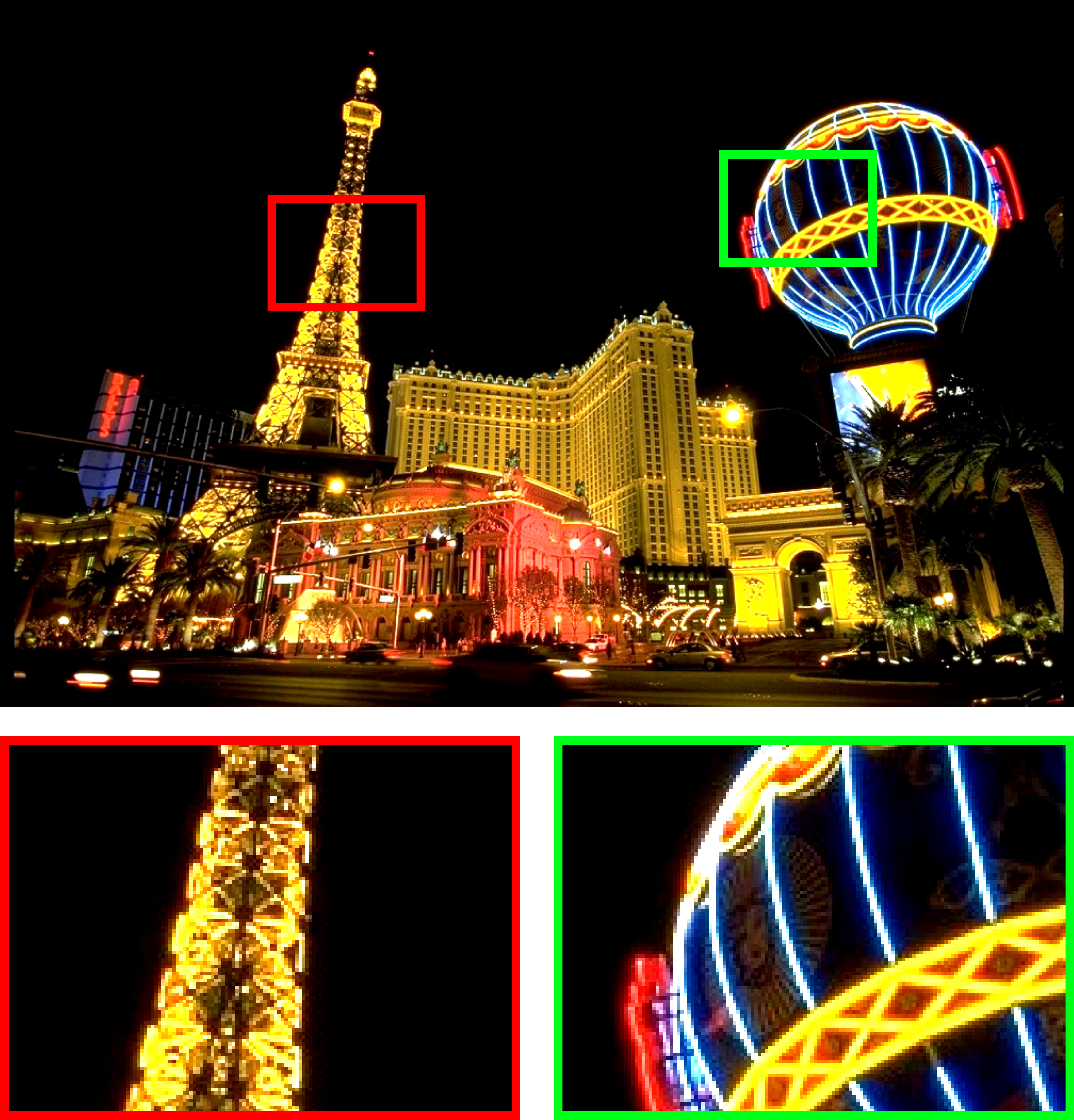}
        \caption{Ground truth}
    \end{subfigure}
    
    \caption{Visual comparison of the deblurring results on the dataset of Lai \textit{et al.} \cite{lai_dataset}. Methods marked with $^*$ are \textit{self-supervised}.}
    \label{fig:lai_overview}
\end{figure*}

\begin{figure*}[t]
    \captionsetup[subfigure]{font=scriptsize,labelfont=scriptsize,{justification=centering}}
    \centering
    \begin{subfigure}[t]{0.15\textwidth}
        \centering
        \includegraphics[width=\linewidth]{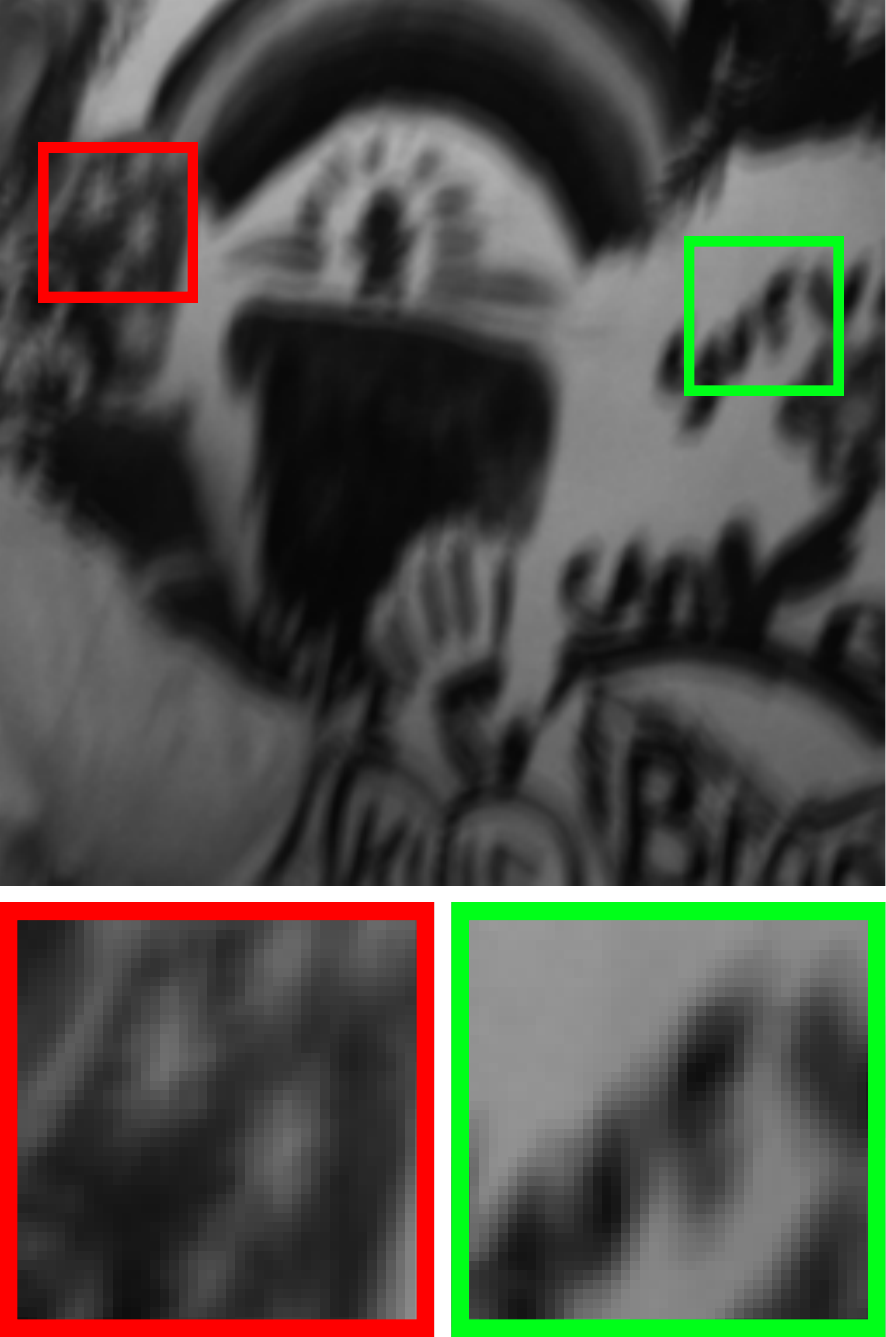}
        \caption{Blurry input}
    \end{subfigure}
    \begin{subfigure}[t]{0.15\textwidth}
        \centering
        \includegraphics[width=\linewidth]{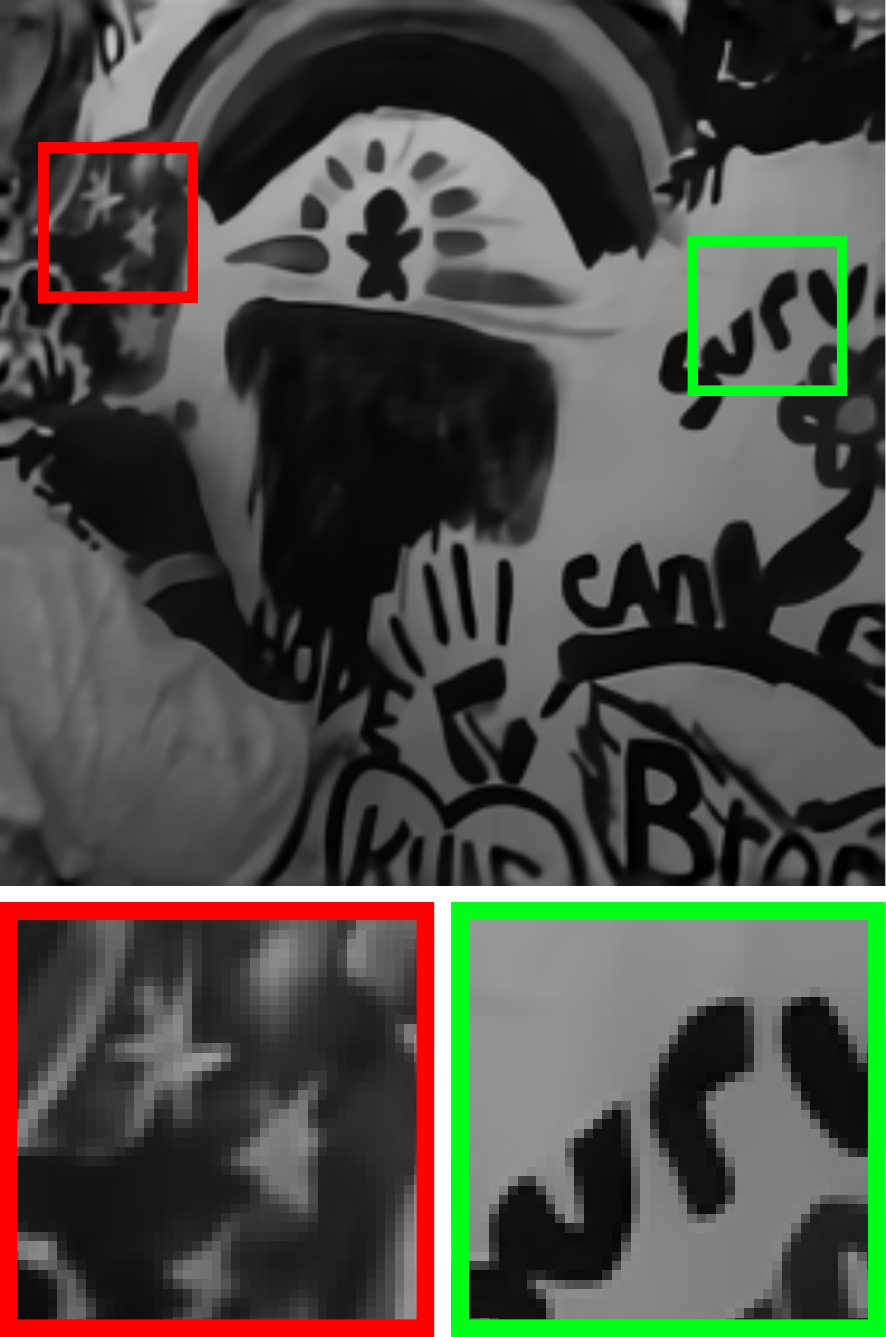}
        \caption{IRCNN\\\cite{IRCNN}}
    \end{subfigure}
    \begin{subfigure}[t]{0.15\textwidth}
        \centering
        \includegraphics[width=\linewidth]{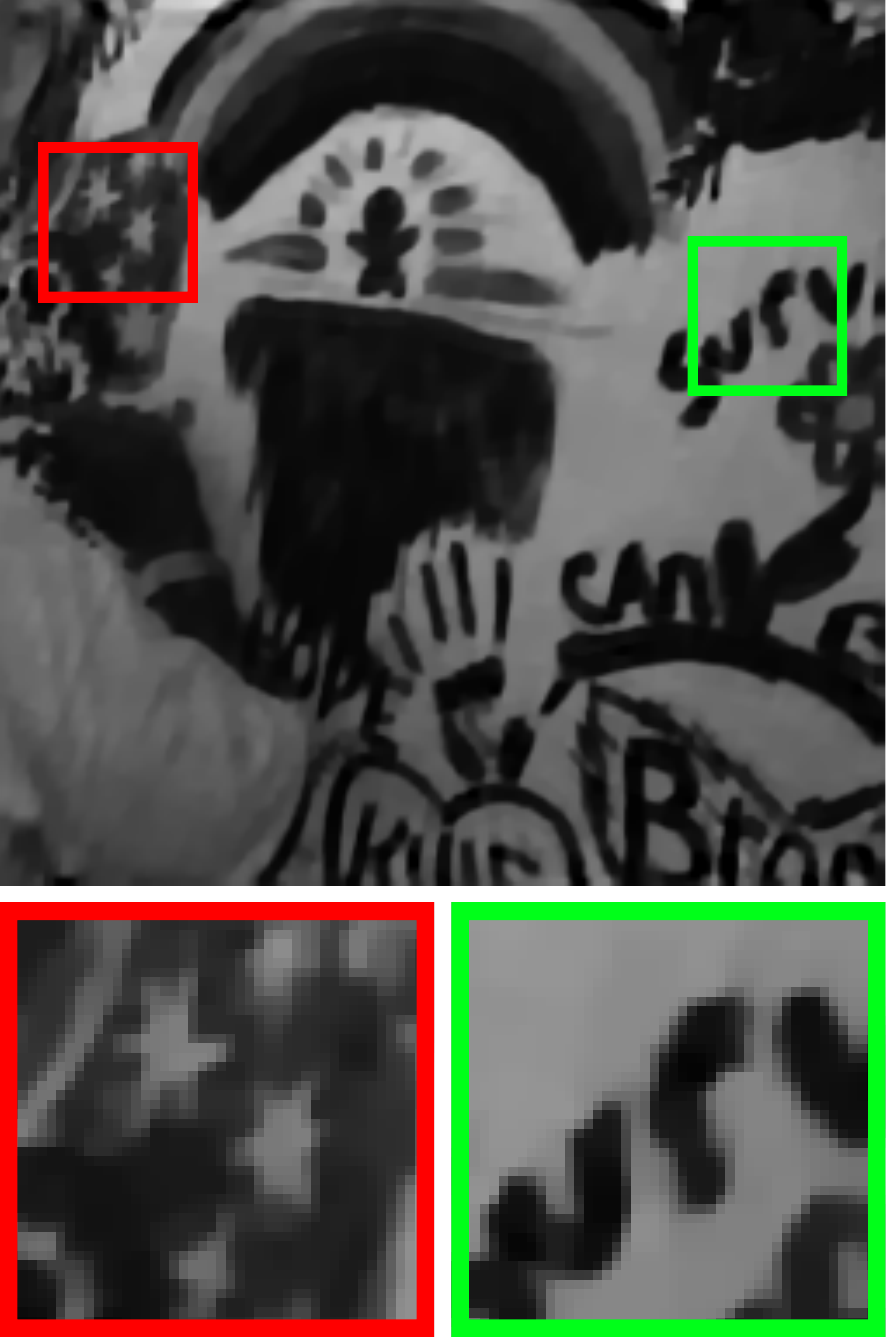}
        \caption{Pan-DCP\\\cite{pan2017deblurring}}
    \end{subfigure}
    \begin{subfigure}[t]{0.15\textwidth}
        \centering
        \includegraphics[width=\linewidth]{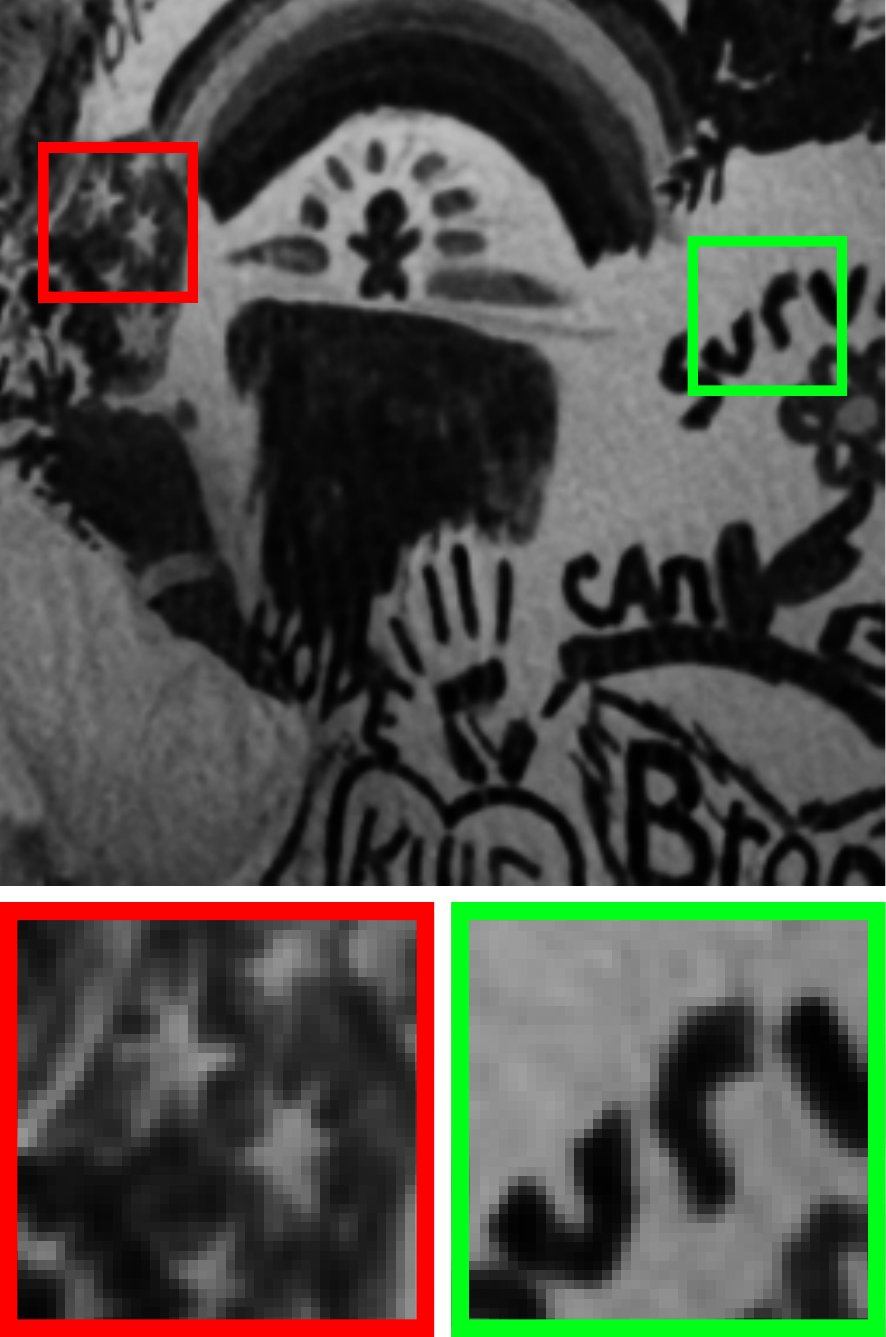}
        \caption{SelfDeblur$^*$\\\cite{selfdeblur}}
    \end{subfigure}
    \begin{subfigure}[t]{0.15\textwidth}
        \centering
        \includegraphics[width=\linewidth]{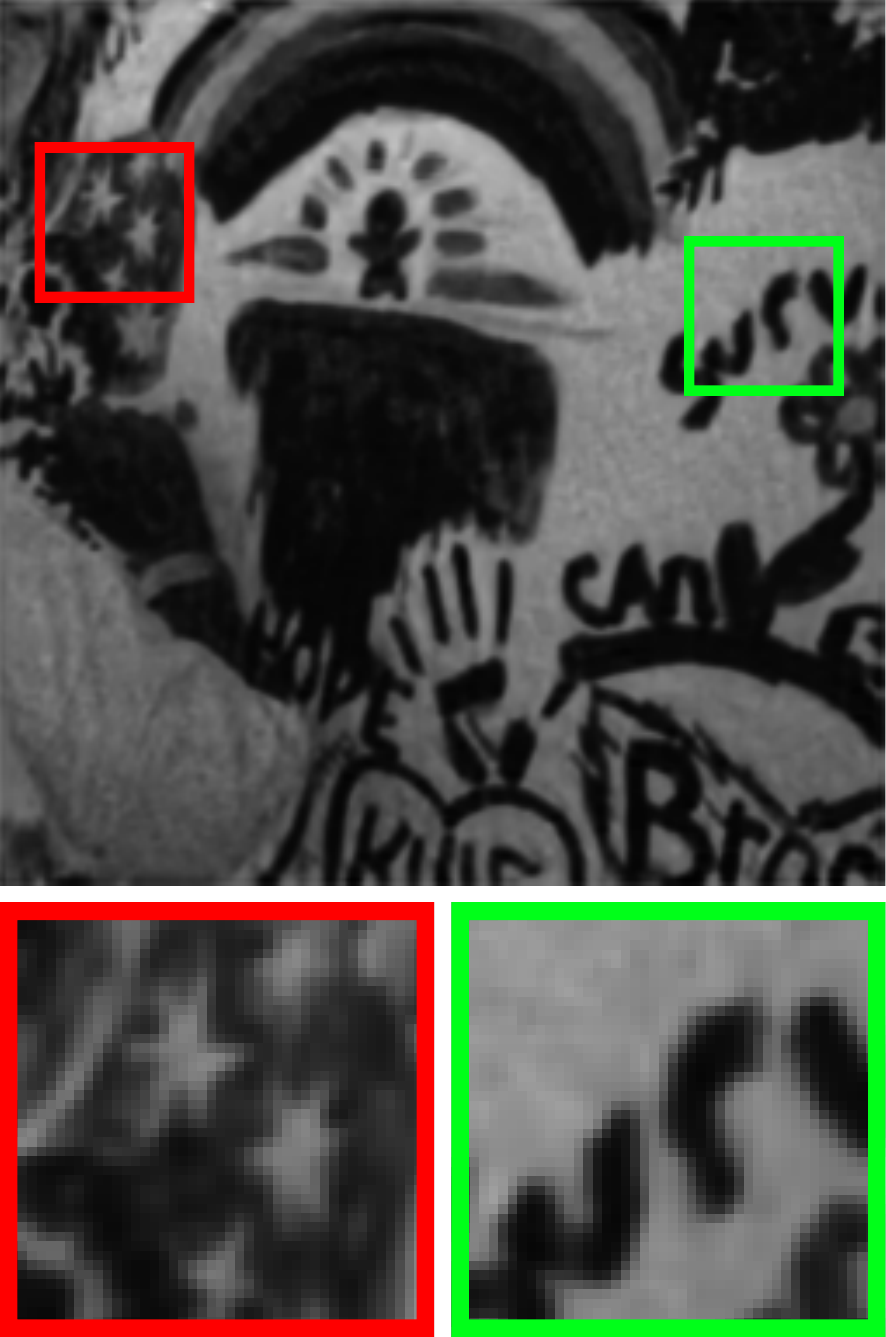}
        \caption{DELAD$^*$}
    \end{subfigure}
    \begin{subfigure}[t]{0.15\textwidth}
        \centering
        \includegraphics[width=\linewidth]{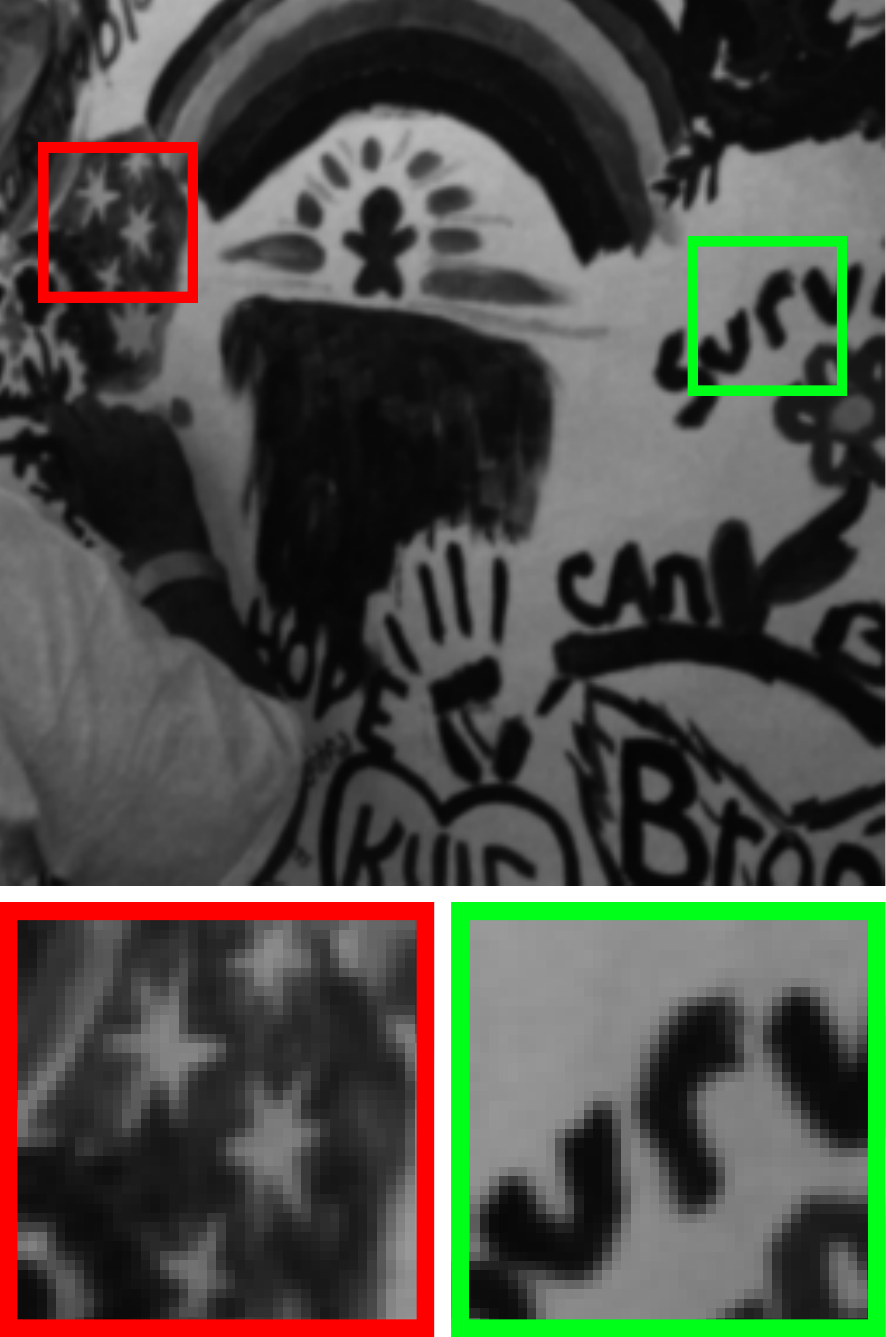}
        \caption{Ground truth}
    \end{subfigure}
    
    \caption{Visual comparison of the deblurring results on the dataset of Levin \textit{et al.} \cite{levin}. Methods marked with $^*$ are \textit{self-supervised}.}
    \label{fig:levin_overview}
\end{figure*}

\subsection{Results on testing images}

\paragraph{Results on the dataset of Levin \textit{et al.} \cite{levin}.} We evaluate our proposed method on the dataset of Levin \textit{et al.} \cite{levin} and compare the performance to other state-of-the-art (SOTA) deconvolution algorithms. Besides deep learning based deconvolution methods, we also include a few classic iterative deconvolution methods like Landweber and Richardson-Lucy methods \cite{landweber, Richardson:72} in our comparison. Table \ref{tab:levin} shows the average PSNR and SSIM values and Figure \ref{fig:levin_overview} shows a few exemplary visual comparisons. Firstly, our method significantly outperforms the classic iterative deconvolution algorithms, highlighting the efficiency and efficacy of embedding the iterative formula inside a deep learning application. Secondly, we are able to obtain better results than SelfDeblur \cite{selfdeblur} in blind and non-blind settings, another SOTA self-supervised method, while having less than 11\% of the number of parameters of non-blind SelfDeblur (without the kernel generator) \cite{selfdeblur}. Lastly, we show in Table \ref{tab:levin} the results obtained with our method without the use of the Hessian in the loss function denoted as DELAD$^{-\mathcal{R}}$. While the performance is still high, the SSIM stays behind non-blind SelfDeblur \cite{selfdeblur}, showing the necessity of inclusion of the Hessian in the loss function for obtaining better performance.

\begin{figure}[]
    \centering
    \includegraphics[width=.8\linewidth]{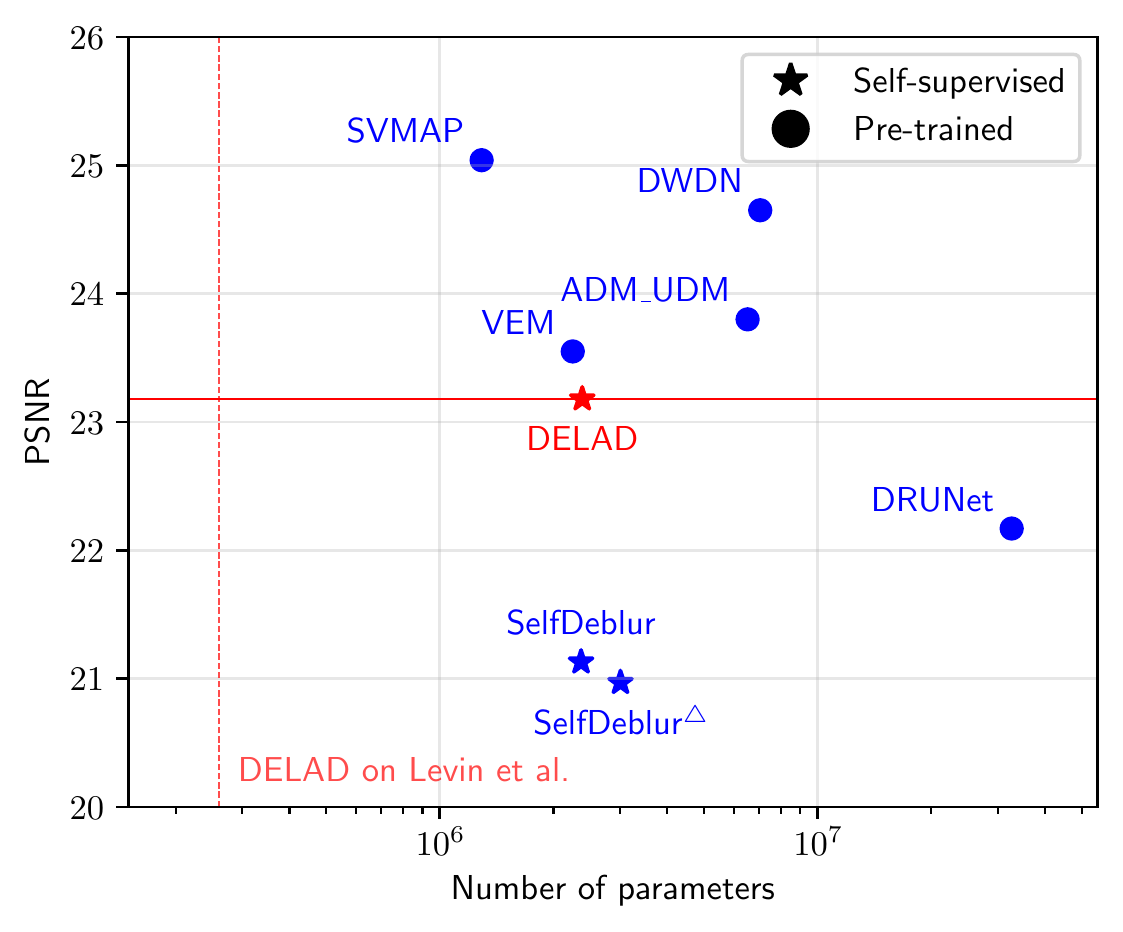}
    \caption{Overview of deconvolution results of state-of-the-art methods (see Table \ref{tab:lai_comparison}) on the dataset of Lai \textit{et al.} \cite{lai_dataset} measured by PSNR ($y$-axis) and their number of parameters ($x$-axis). Notice the $x$-axis being on a logarithmic scale. $^\triangle$ indicates the deblurring results were obtained using a blind deconvolution setting. We include the number parameters required for deconvolving images from the dataset of Levin \textit{et al.} \cite{levin} with DELAD as well. Because the terms $\mathbf{x}^{(0)},\mathbf{m}^{(1)},\mathbf{m}^{(2)},\mathbf{m}^{(3)}$ are learnable matrices of the same size as the input image, the number of model parameters changes based on the size of the input.}
    \label{fig:parameters_psnr}
\end{figure}

\paragraph{Results on the dataset of Lai \textit{et al.} \cite{lai_dataset}.} While the Levin \textit{et al.} \cite{levin} dataset consists of grayscale images, the dataset of Lai \textit{et al.} \cite{lai_dataset} contains color images. We used the same approach as Ren \textit{et al.} \cite{selfdeblur} to perform the deconvolution by splitting the images into the YCbCr channels and only doing the optimisation for the Y channel. The average PSNR and SSIM values are shown in Table \ref{tab:lai_comparison} and the visual comparison is in Figure \ref{fig:lai_overview}. While SOTA methods that use large pre-trained networks can obtain high-quality reconstructed images, our self-supervised method is still able to obtain very competitive results and even outperform some of the pre-trained networks. Figure \ref{fig:parameters_psnr} highlights the performance of the model. From the visual comparison in Figure \ref{fig:lai_overview}, it can be seen that the inclusion of Hessian prior allows for much smoother structures particularly apparent in the background. Moreover, the continuity assumption in the loss function leads to more polished edges compared to other methods.

\paragraph{Results on microscopy dataset \cite{wiener-dip}.} We also evaluate DELAD on simulated microscopy images and show its capacity to deconvolve images similar to a real scenario with a strong blur expected in two-photon microscopy. The quantitative results and comparison with classic iterative algorithms used in deconvolution of microscopy images as well as self-supervised methods are shown in Table \ref{tab:wiener-dip}. Because the simulated images have natural grain in the background, which breaks out the assumption of continuity, we also evaluate our model without the use of the Hessian in the loss function (denoted as DELAD$^{-\mathcal{R}}$). Indeed, the Hessian regularization naturally suppresses such grain in the deconvolved images, therefore DELAD$^{-\mathcal{R}}$ achieves better results than DELAD. Both versions of our algorithm outperform other self-supervised deconvolution methods.

\begin{table*}[t]
\caption{Evaluation of the average PSNR and SSIM of the deblurring results on every class from the dataset of Lai \textit{et al.} \cite{lai_dataset}. We highlight the \textbf{highest} and the \underline{\textit{lowest}} values for each class. Note that while our method is completely \textit{self-supervised} it still achieves very competitive results compared to state-of-the-art methods that leverage large pre-trained networks.}
\vspace{2mm}
\resizebox{\textwidth}{!}{
\begin{tabular}{llcccccccc}
\toprule
\multicolumn{1}{l}{Class} &           & Pan-DCP & IRCNN & ADM\_UDM & DRUNet & VEM & DWDN & SVMAP & \textbf{DELAD} \\
\multicolumn{1}{l}{} &           & \cite{pan2017deblurring} & \cite{IRCNN} & \cite{ADMUDM} & \cite{DRUNet} & \cite{VEM} & \cite{dwdn} & \cite{dong2021learning} & \\
\midrule
\multirow{2}{*}{\textsc{Manmade}}            & PSNR      & \underline{\textit{18.59}} & 20.47 & 22.43 & 20.62 & 22.71 & \textbf{24.02} & 23.75 & 21.94 \\
                                             & SSIM      & \underline{\textit{0.594}} & 0.604 & 0.724 & 0.613 & 0.780 & \textbf{0.836} & 0.776 & 0.628 \\
\midrule
\multirow{2}{*}{\textsc{Natural}}            & PSNR      & \underline{\textit{22.60}} & 23.26 & 25.04 & 23.25 & 25.29 & 25.91 & \textbf{26.23} & 24.84 \\
                                             & SSIM      & 0.698 & 0.636 & 0.733 & \underline{\textit{0.630}} & 0.752 & \textbf{0.814} & 0.778 & 0.675 \\
\midrule
\multirow{2}{*}{\textsc{People}}             & PSNR      & \underline{\textit{24.03}} & 28.04 & 28.81 & 28.04 & 27.19 & 30.02 & \textbf{30.88} & 27.63 \\
                                             & SSIM      & 0.772 & 0.843 & 0.866 & 0.838 & \underline{\textit{0.723}} & \textbf{0.905} & 0.899 & 0.830 \\
\midrule
\multirow{2}{*}{\textsc{Saturated}}          & PSNR      & \underline{\textit{16.52}} & 16.99 & 17.57 & 17.14 & 17.65 & 17.90 & 18.75 & \textbf{18.91} \\
                                             & SSIM      & 0.632 & 0.642 & 0.627 & 0.658 & \underline{\textit{0.600}} & 0.695 & \textbf{0.733} & 0.666 \\
\midrule
\multirow{2}{*}{\textsc{Text}}               & PSNR      & \underline{\textit{17.42}} & 21.37 & 25.13 & 21.79 & 24.92 & 25.40 & \textbf{25.60} & 22.56 \\
                                             & SSIM      & \underline{\textit{0.619}} & 0.828 & 0.883 & 0.829 & 0.853 & 0.877 & \textbf{0.894} & 0.836 \\
\midrule
\multirow{2}{*}{Overall}                     & PSNR      & \underline{\textit{19.89}} & 22.03 & 23.80 & 22.17 & 23.55 & 24.65 & \textbf{25.04} & 23.18 \\
                                             & SSIM      & \underline{\textit{0.666}} & 0.710 & 0.767 & 0.714 & 0.742 & \textbf{0.825} & 0.816 & 0.727 \\
\bottomrule
\label{tab:lai_comparison}
\end{tabular}}
\end{table*}

\begin{table}[]
    \caption{Comparison of the average PSNR and SSIM of the deblurring results on the dataset of Levin \textit{et al.} \cite{levin} using iteration-based self-supervised deconvolution methods. $^\triangle$ indicates the deblurring results were obtained using a blind deconvolution setting. DELAD$^{-\mathcal{R}}$ denotes our method without the Hessian in its loss function.}
    \label{tab:levin}
    \vspace{2mm}
    \centering
    \begin{tabular}{lcc}
        \toprule
        Method   & PSNR (dB) & SSIM  \\
        \midrule
        LW \cite{landweber}        & 18.02     & 0.568 \\
        RL \cite{Richardson:72}        & 18.71     & 0.630 \\
        $\text{D-URL}^\triangle$ \cite{deep-url}      & 27.12     & 0.910 \\
        IRCNN \cite{IRCNN} & 31.89 & 0.933 \\
        Pan-DCP \cite{pan2017deblurring}        & 32.69     & 0.928 \\
        $\text{SelfDeblur}^\triangle$ \cite{selfdeblur} & 33.07     & 0.931 \\
        SelfDeblur \cite{selfdeblur} & 33.32     & 0.944 \\
        \midrule
        DELAD$^{-\mathcal{R}}$       & 34.07     & 0.938 \\
        DELAD       & \textbf{34.35}     & \textbf{0.954} \\
        \bottomrule
    \end{tabular}
\end{table}

\begin{table}[]
    \centering
    \caption{Comparison of the average PSNR and SSIM of the deblurring results on the microscopy dataset  \cite{wiener-dip} using iteration-based self-supervised deconvolution methods. $^\triangle$ indicates the deblurring results were obtained using a blind deconvolution setting. DELAD$^{-\mathcal{R}}$ denotes our method without the Hessian in its loss function.}
    \label{tab:wiener-dip}
    \vspace{2mm}
    \begin{tabular}{lcc}
        \toprule
        Method   & PSNR (dB) & SSIM \\
        \midrule
        LW \cite{landweber}         &   9.57 &  0.166 \\
        RL \cite{Richardson:72}        &   9.91   & 0.180 \\
        $\text{W-DIP}^\triangle$ \cite{wiener-dip} & 22.12     & 0.419 \\
        SelfDeblur \cite{selfdeblur} & 22.44 & 0.443 \\
        \midrule
        DELAD$^{-\mathcal{R}}$ &  \textbf{23.70}  &  \textbf{0.640}  \\
        DELAD       &  22.88  &  0.482  \\
        \bottomrule
    \end{tabular}
\end{table}

% \begin{figure}[t]
%     \centering
%     \includegraphics[width=.7\linewidth]{figures/microscopy1/microscopy-01.pdf}
%     \caption{Example of deconvolution of synthetic microscopy images from simulation of Schneider \textit{et al.} \cite{SCHNEIDER2015220} blurred with PSFs expected in two-photon microscopy, provided by Bredell \textit{et al.} \cite{wiener-dip}. The input images blurred with various kernels are shown in column (a), (b) shows the deconvolution results with DELAD$^{-\mathcal{R}}$, and (c) are the ground truth sharp images.}
%     \label{fig:microscopy_overview}
% \end{figure}

\subsection{Results on real EDOF microscopy images}

\begin{figure*}[t]
    \centering
    \includegraphics[width=\linewidth]{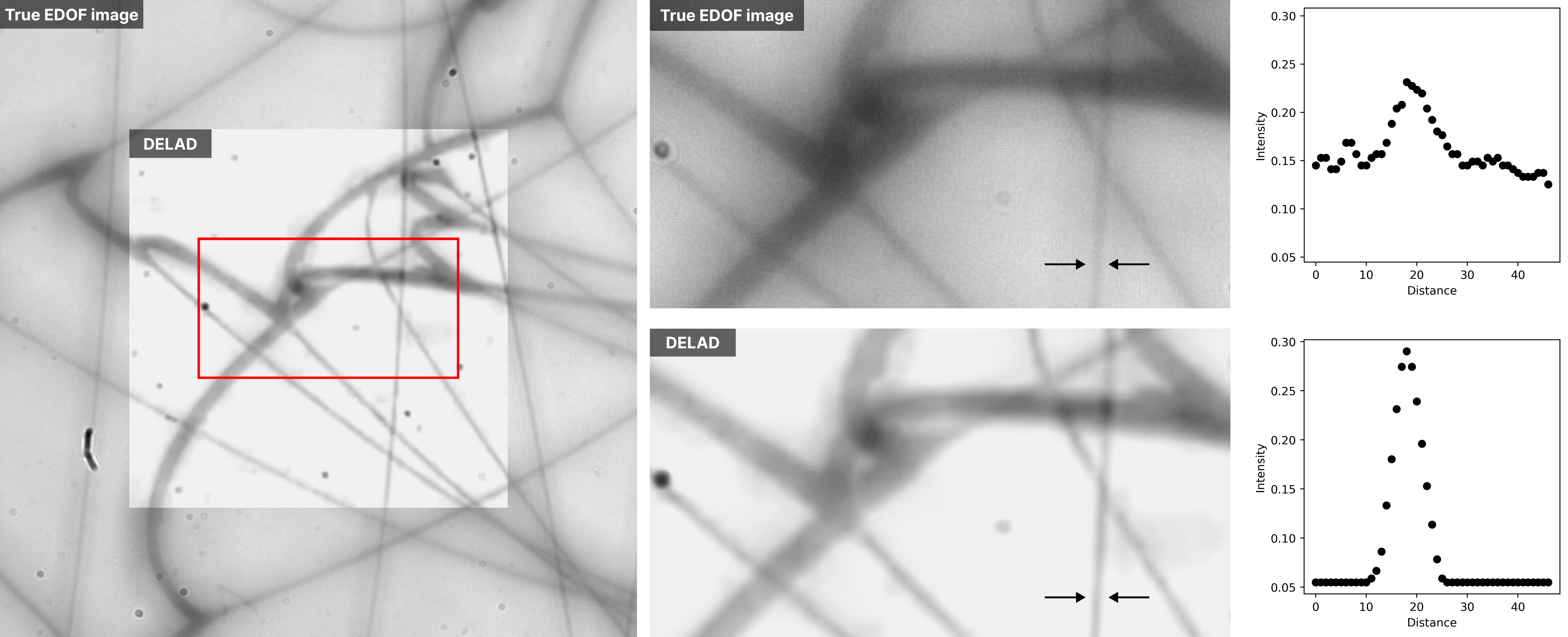}
    \caption{Overview of the sparse deconvolution of real EDOF image obtained by our underwater imaging system. The inverse pixel intensity plots on the right show the background suppression and contrast increase in the deconvolved image compared to the input, which is crucial for subsequent classification and segmentation tasks.}
    \label{fig:edof-large}
\end{figure*}

\begin{figure*}[h!]
    \centering
    \includegraphics[width=\linewidth]{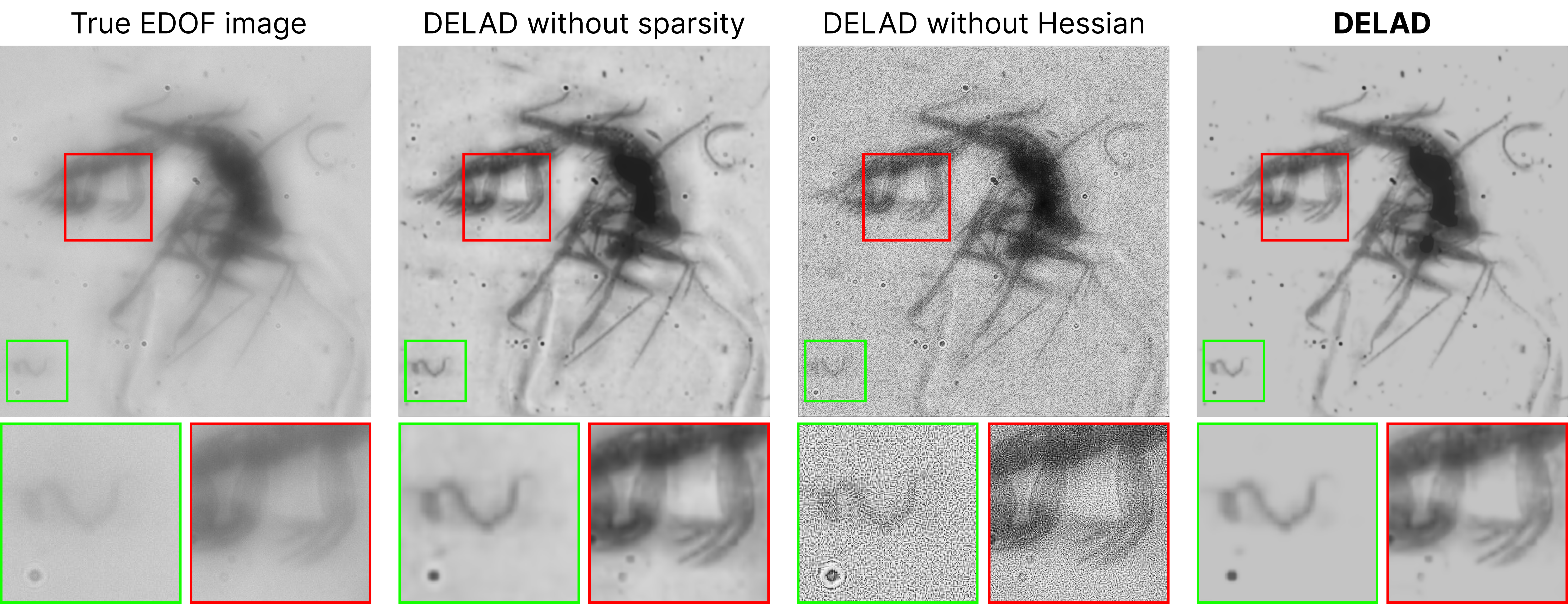}
    \caption{Demonstration of the necessity of Hessian and the sparsity constrain in the loss function. Without the Hessian, the resulting image is noisy with reduced visibility of the fine details and structures in the image. Sparsity smooths the background and improves the overall legibility.}
    \label{fig:edof-small-1}
\end{figure*}

We then apply our method to real extended-depth-of-field (EDOF) images. We deconvolve all EDOF images with the same kernel simulated following \cite{Liu:11} with the Gibson-Lanni model \cite{Gibson:92}. Before the deconvolution step, we estimate and remove the background as described above as a pre-processing step. The background-removed image is then deconvolved using our method that is optimised in 1000 epochs. The initial learning rate is set to $5\mathrm{e}{-3}$ with a decaying factor of $0.2$ when reaching epoch 700. During optimisation, Equation \ref{eq:loss_sparse} is being minimised, where $\psi_1=3\mathrm{e}{-6}$ and $\psi_2=0.2$. The deconvolution results in Figure \ref{fig:edof-large} show increased contrast and suppression of the haze surrounding the foreground objects, one characteristic blurry pattern in EDOF images. We also plot inverse pixel intensity to show increased contrast and background suppression with our method. Figure \ref{fig:edof-small-1} further shows the difference in deconvolution performance between DELAD, DELAD$^{-\mathcal{R}}$ (omitting the Hessian from the loss function) and DELAD optimised without the sparsity constraint. Without the Hessian in the loss function, the reconstructed image is noisy and lacks the clarity of the structures and fine details in the image. The inclusion of the Hessian regularization is essential for obtaining smooth recovered objects. Besides Hessian, the sparsity regularization can further elevate the quality of the resulting deconvolved image. The results obtained with DELAD without the sparsity prior in the loss function include noisy patches as the EDOF haze blurry pattern is not completely suppressed, hence demonstrating itself to be an essential part of the algorithm when deconvolving the real EDOF microscopy images. The results shown in Figure \ref{fig:edof-small-1} highlight the strength of DELAD to increase the contrast, optical resolution, and legibility of the objects on the deconvolved images whereas in the raw EDOF images the animal structures are hardly visible. The improved images thereby facilitate downstream processing (e.g. detection and segmentation) as well as taxonomic classification.

% maybe show (in the supplementary) how DELAD outperforms other algorithms on our EDOF images, e.g. the ones you also show in Fig 3 and 4

% \begin{table}[t]
%     \centering
%     \includegraphics[width=.85\linewidth]{figures/edof/edof-grid-2.pdf}
%     \caption{Overview of DELAD deconvolution results of a microscopy resolution test target with different loss functions. The Hessian and sparsity prior in the loss function results in clearest and most legible reconstructions.}
%     \label{fig:edof-small-2}
% \end{table}

% ------------------------------------------------------------------------
\section{Conclusion}

In this paper, we propose a lightweight self-supervised non-blind deconvolution module that embeds a classic iterative deconvolution algorithm into a deep learning application. By integrating convolutional layers into the iterative Landweber deconvolution algorithm we are able to efficiently reconstruct the sharp image. We also explore constraints added to the data fidelity term, Hessian and sparse prior. We show that our model is able to obtain very competitive quantitative results when evaluated on computer vision benchmark datasets, without leveraging any large pre-trained networks and having only a fraction of parameters compared to other SOTA methods. Moreover, we evaluate our method on real microscopy images obtained by our EDOF underwater imaging system showing the capabilities of our model in a real-world application.

% ------------------------------------------------------------------------
%%%%%% REFERENCES
% \bibliographystyle{plain} % We choose the "plain" reference style
\bibliography{aaai23.bib} % Entries are in the refs.bib file

\end{document}